\documentclass[aps,twocolumn,showpacs]{revtex4}

\RequirePackage{amsfonts,amssymb,amsmath}

\begin{document}
\title{ Quantum singularities in the BTZ spacetime}
\author{Jo\~ao Paulo M. Pitelli} 
\email{e-mail:pitelli@ime.unicamp.br}
\author{Patricio S. Letelier} 
 \email{e-mail: letelier@ime.unicamp.br} 
\affiliation{
Departamento de Matem\'atica Aplicada-IMECC,
Universidade Estadual de Campinas,
13081-970 Campinas,  S.P., Brazil}


\begin{abstract}

The spinless Ba\~nados-Teiltelboim-Zanelli (BTZ) spacetime is considered 
 in the quantum theory  context. Specially, we study the case of  negative
 mass parameter  using  quantum test particles  obeying the  Klein-Gordon 
and Dirac equations. We  study  if this classical singular  spacetime, with
 a naked singularity at the origin,  remains singular when tested with 
 quantum particles. The need of additional information near the   origin 
is confirmed for massive scalar particles  and all the possible boundary 
conditions necessary to turn the spatial portion  of the wave operator  
 self-adjoint are found.  When tested by massless scalar particles or 
fermions, the singularity is ``healed'' and no  extra boundary condition 
are needed. Near infinity, no boundary conditions are necessary.

\end{abstract}


\maketitle


\section{INTRODUCTION}
\label{INTRODUCTION}

In the  $2+1$ dimensional Einstein theory of gravitation without cosmological constant, the spacetime 
 is necessarily  flat and only its global topological properties \cite{jackiw1} makes it different 
 from the trivial $2+1$ dimensional Minkowski spacetime. In the simplest case of a point particle at 
 the origin, the resulting spacetime is conic \cite{staruszkiewicz} (the usual plane with a slice 
 removed and identified edges). It is the $2+1$ dimensional analog of the $3+1$ dimensional cosmic 
 string (for a classical treatment see \cite{letelier} and for a quantum treatment
 \cite{konkowski1}). 

When a negative cosmological constant, $\Lambda$,  is considered, the Einstein equations 
admit a  black hole solution \cite{btz1}. The lower dimension of the BTZ solution makes it 
 a particularly  simple example of a spacetime with the main  properties of the
 usual $3+1$ black hole.

The negative cosmological constant gives us a   asymptotically anti-de Sitter spacetime, 
instead of a flat one. In fact, the BTZ spacetime is locally
 anti-de Sitter, differing only by its  global topological properties \cite{btz2,jackiw2}.

There are three different kinds of spacetimes (solutions of Einstein equations) 
depending  on a mass parameter  $m$, which has been adjusted so that the mass 
vanishes when the horizon size goes to zero, that is, $m=0$ for the vacuum state. 
For $m > 0$, there is a continuum black hole spectrum with a singularity of  the 
 Taub-NUT type at the origin, hidden by an event horizon given by 
$r_{+}=\sqrt{m}l$ \cite{btz1},  where  $l^{-2}=-\Lambda$. This spacetime  
does not violate the cosmic censorship hypothesis since the singularity 
  is hidden. It is a reasonable classical spacetime and   quantum mechanical
considerations are not needed in this case.
 
  As $m$ takes values  smaller than or equal to zero, there appears a continuous
 sequence of naked singularities  (point particle sources) at the origin. The 
singularities  do  not come from  any curvature scalar 
  divergence, but rather from a topological obstruction of the spacetime continuation, 
since the  Ricci tensor   has a term proportional to the Dirac distribution 
\cite{pantoja} in addition to the constant curvature. This last  term is  due to 
the presence
 of the cosmological constant.  Near the  origin, where the curvature can be 
 neglected, the spacetime is conic, so  it must be excluded by the cosmic 
censorship hypothesis.  It is in this classical background that the quantum test
 particles will be studied in order to see if the  spacetime remains singular
 when considered in the  quantum theory context.  In this paper we adopt the definition 
 of quantum singularity due to Horowitz and Marolf \cite{horowitz}, which says 
that a spacetime is  quantum mechanically nonsingular if the time evolution of 
any wave packet is uniquelly determined by  the initial wave function. 
 
 When $m$ takes the value $-1$, the spacetime does not present an event horizon, 
but there is no singularity to hide either, so  this solution (a true anti-de 
Sitter spacetime) is  again permissible and it is the ground state of the 
 theory (for a discussion of the importance of naked singularities  to establish 
the ground state in any  gravitation theory, see \cite{horowitz2}). 
 
 For the mass parameter \ $m<-1$  the spacetime represents point sources 
with negative mass without physical meaning.

The purpose of this work is to study the naked singularities for the continuous
 sequences of spacetimes  separating the black hole like spectrum from the ground 
state anti-de Sitter spacetime. We shall use  quantum test particles to determine 
 if these spacetimes  are quantum mechanically singular.

The paper is organized as follows, in section \ref{QUANTUM SINGULARITIES} we
 present a brief review of  quantum singularities in a general static spacetime. 
In section \ref{SCALAR} we apply the formalism  presented in the previous section 
 to the case of scalar particles.  We also consider the boundary conditions 
studied by Kay and Studer \cite{kay} and adopt them in the  context of the  BTZ 
 spacetime. In section  \ref{SPINOR}, we extend the formalism to  particles with
  spin. Finally, in section \ref{CONCLUSION}{\bf ,} we  discuss the results 
presented in this work.


\section{QUANTUM SINGULARITIES} 
\label{QUANTUM SINGULARITIES}

In general relativity, a spacetime singularity is indicated by incomplete 
geodesics or, more precisely, by b-incompleteness, i.e., incomplete curves of 
boundary acceleration \cite{hawking}. At the singular points, an extra information 
must be added,  since we lose the capacity to predict the future of a particle 
following an incomplete worldline.

In order to generalize this concept to quantum mechanics, Horowitz and Marolf
 proposed a simple definition of singularity. They stated that a spacetime is
 nonsingular if the evolution of any state is  uniquely determined for all time.
In this way, a spacetime is said quantum mechanically nonsingular if the time 
 evolution of any wave packet is uniquely determined by the initial wave data 
on a Cauchy surface.

To be more precise,  let $(M,g_{\mu\nu})$ be a static spacetime with a 
timelike Killing vector field $\xi^{\mu}$, $t$ be the Killing parameter 
 and $\Sigma$ a static spatial slice orthogonal to $\xi^{\mu}$. 
 The Klein-Gordon equation on this spacetime,
\begin{equation}
\square\Psi=M^2\Psi ,
\label{Klein-Gordon_0}
\end{equation}
can be split in a temporal and a spatial part, 
\begin{eqnarray}
\frac{\partial^2 \Psi}{\partial t^2}=-A\Psi=VD^{i}(VD_{i}\Psi)+M^2 V^2 \Psi,
\label{separada}
\end{eqnarray}
where $V^2=-\xi^{\mu}\xi_{\mu}$ and $D_{i}$ is the spatial covariant 
derivative  on $\Sigma$

To avoid the singular points, we take $C_{0}^{\infty}(\Sigma)$, the set of all 
smooth functions of compact support on $\Sigma$,  as the domain $D(A)$ of the 
operator $A$ defined in equation (\ref{separada}). With this domain, $A$ is a 
well-defined positive  symmetric operator on the Hilbert space $H=L^2(\Sigma,
 V^{-1}d\mu)$, where $d\mu$ is the usual measure on $\Sigma$. 
 
 The chosen  domain is so small, i.e., the restrictions on functions are so 
strong, that the domain of the Hilbert adjoint operator $A^{\ast}$ 
 is extremely large and it is composed of all functions
 $\psi$ in $L^2(\Sigma,V^{-1}d\mu)$ such that $A\psi \in L^2$. Then, $A$ 
 is not self-adjoint. 
 
 Hence, we are face with the problem of searching for self-adjoint extensions 
of $A$ and to discover if it has only   one or many of such extensions.

If $A$ has only one self-adjoint extension (its closure $\overline{A}$), then
 $A$ is said essentially self-adjoint
 \cite{reed-simon1,reed-simon2,richtmyer}. Since  we are worried with  the 
one particle description, not a field theory,  the positive frequency solution satisfies
\begin{equation}
i\frac{\partial \Psi}{\partial t}=(\overline{A})^{1/2}\Psi,
\end{equation}  
and the evolution of the  wave packet is uniquely determined by the 
initial data,
\begin{equation}
\Psi(t,{\bf x})=e^{-it(\overline{A})^{1/2}}\Psi(0, {\bf x}).
\end{equation}
In this case, we say  that the spacetime is quantum mechanically non-singular.

Now, if $A$ has many self-adjoint extensions $A_{\alpha}$, where $\alpha$ is a
 real parameter, we must choose one  in order to evolve the wave packet. Any 
solution of the form
\begin{equation}
\Psi(t, {\bf x})=e^{-it(A_{\alpha})^{1/2}}\Psi(0, {\bf x}),
\end{equation} 
is a good solution and an extra information must be given to tell us which one
 has to be chosen. In this case we say that the  spacetime 
 is quantum mechanically singular.

The criterion used to determine the number of self-adjoint extensions of
 $A$ (Theorem $X.2$ on reference  \cite{reed-simon2}) is to solve the equations
\begin{equation}
A^{\ast}\psi \pm i\psi=0, 
\label{teste}
\end{equation}
and  to count the number of linear independent solutions in $L^2$, i.e., 
the dimension of $\ker(A^{\ast}\pm i)$. 
 If there is no  square-integrable solutions, the operator posses a unique 
self-adjoint extension and it is essentially self-adjoint. If there is one 
solution in   $L^2$ to each equation in (\ref{teste}), a one-parameter family 
of self-adjoint extensions exists and its extension is not unique.
 The theory of deficiency indices of von Neumann says that these self-adjoint
 extensions are represented by the one-parameter family of the extended 
domains of the operator $A$ given by \cite{reed-simon2}
 \begin{equation}
D^{\omega}=\{\psi=\phi +\phi^{+}+e^{i\omega}\phi^{-}:\omega \in \mathbb{R},\,\phi \in D(A)\},
\label{neumann}
\end{equation}
where
\begin{equation}
A^{\ast}\phi^{\pm}=\pm i \phi^{\pm}
\end{equation}
and $\phi^{\pm}\in L^2$. The term $e^{i\omega}$ in (\ref{neumann}) appears because 
the theory says that the self-adjoint  extensions of the operator $A$ are in 
one-to-one correspondence with the isometries from $\ker(A^{\ast}-i)$ to 
 $\ker(A^{\ast}+i)$, i.e., the isometries given by $\phi^{+}\mapsto e^{i\omega}\phi^{-}$. 
 
\section{SCALAR FIELDS}
\label{SCALAR}

The metric for the spinless BTZ   spacetime \cite{btz1}  is 
\begin{equation}
ds^2=-V(r)^2dt^2+V(r)^{-2}dr^2+r^2d\theta^2,
\end{equation}
with the usual ranges of the  cylindrical coordinates and 
\begin{equation}
V(r)^2=-m+\frac{r^2}{l^2},
\end{equation}
where $m$ is the mass parameter.

After separating variables, $\psi=R(r)e^{in\theta}$, the radial portion 
of equation (\ref{teste}) can be cast as
\begin{equation}
R_{n}''+\frac{(V^2 r)'}{V^2 r}R_{n}'-\frac{n^2}{V^2r^2}R_{n}
-\frac{M^2}{V^2}R_{n}\pm i\frac{R_{n}}{V^4}=0.
\label{eqhorowitz}
\end{equation}

To consider the case $r\to\infty$, we note that for large values of $r$
 the metric takes the form,
\begin{equation}
ds^2\approx -\bigg(\frac{r^2}{l^2}\bigg)dt^2+\bigg(\frac{r^2}{l^2}
\bigg)^{-1}dr^2+r^2d\theta^2.
\label{infinito}
\end{equation}
This spacetime is asymptotically anti-de Sitter. The measure on the 
slice $\Sigma$ is $d\mu=rdr$. From the  definition of Horowitz and 
Marolf \cite{horowitz} we find that the measure of our Hilbert space
 $H$ is $V^{-1}d\mu=ldr$.

Then, the equation (\ref{eqhorowitz}) takes the form
\begin{equation}
R_{n}''+\frac{3}{r}R_{n}'-\frac{n^2l^2}{r^4}R_{n}
-\frac{M^2l^2}{r^2}R_{n} \pm i\frac{l^4R_{n}}{r^4}=0.
\label{eqhorowitz2}
\end{equation}

For very large values of $r$, we can consider 
only the first two terms of the equation (\ref{eqhorowitz2}). Then we have
\begin{equation}
R_{n}''+\frac{3}{r}R_{n}'=0,
\end{equation}
whose solution is
\begin{equation}
R_{n}(r)=C_{1n}+C_{2n}r^{-2},
\end{equation}
where $C_{1n}$ and $C_{2n}$ are arbitrary constants. $R(r)\in L^2$ 
if and only if $C_{1n}=0$. Then, for each mode we have only one solution
 in $L^2$. Let us now, analyze the case $r \to 0$.  

The metric in this case is approximately given by
\begin{equation}
ds^2\approx -\alpha^2dt^2+\alpha^{-2}dr^2+r^2d\theta^2,
\end{equation}
where $\alpha^2=-m$ (remember we are interested in the case $-1<m<0$).

Redefining the coordinates ($t\to\alpha$, $r\to\alpha^{-1}r$), we have
\begin{equation}
ds^2\approx -dt^2+dr^2+\alpha^2r^2d\theta^2.
\label{cosmic}
\end{equation}

The metric (\ref{cosmic}) tells us  that near the singularity $r=0$, 
where curvature effects are negligible, the BTZ spacetime is conic.

The parameter 
$m=-\alpha^2$ is related to the mass of the  point particle source  by 
$\alpha=1-4Gm_{\textrm{source}}$ and to the deficit angle by 
$\Delta=2\pi(1-\alpha)$.

As noted by Horowitz and Marolf \cite{horowitz}, the case of the massive 
test  particles  need not to be considered. The additional 
term $-\frac{M^2}{V^2}R$ in equation (\ref{eqhorowitz}) acts as
 a repulsive potential, increasing the rate at which the non-square
integrable solution diverges at the origin, and driving the square 
integrable solution more quickly to zero. Then,  if the operator $A$ defined 
in equation (\ref{separada}) is essentially self-adjoint for $M=0$, it 
is also essentially self-adjoint for $M>0$. Therefore, we need only  to analyze
the massless case $M=0$.

From  (\ref{cosmic}) we find that  equation (\ref{eqhorowitz}) reduces to

\begin{equation}
R_{n}''+\frac{1}{r}R_{n}'+\bigg[\pm i-\frac{n^2}{\alpha^2r^2}\bigg]R_{n}=0,
\label{cosmic2}
\end{equation}
whose  general solution is
\begin{equation}
R_{n}(r)=A_{n}J_{|n/\alpha|}(kr)+B_{n}N_{|n/\alpha|r},
\label{solution}
\end{equation}
where $J_{\nu}(\kappa r)$ and $N_{\nu}(\kappa r)$ are  the $\nu^{\rm th}$ order Bessel and 
Neumann functions, respectively, and $k=\sqrt{i}$.

Near $r=0$, $J_{|n/\alpha|}(x)\sim x^{|n/\alpha|}$, $\forall \; n$, 
while $N_{|n/\alpha|}(x)\sim x^{|n/\alpha|}$, except for $n=0$, when
 $N_{|n/\alpha|}(x)\sim \ln x$. From the behavior of the Bessel and Neumann
 functions near the origin, it is easy to show that $J_{|n/\alpha|}(kr)$ is
 square-integrable near $r=0$ for all \mbox{$n=0,1,2,\dots$}, 
while $N_{|n/\alpha|r}$ is square-integrable near $r=0$ only 
if $|n/\alpha|\leq 1$, or $|n|\leq \alpha<1$. So $N_{n}(r)$ belongs 
to $L^2$ near $r=0$ only if $n=0$. In this case, we can adjust the
 constants $A_{0}$ and $B_{0}$ in equation (\ref{solution}) to meet 
the asymptotic behavior at infinity, $R(r)\sim 1/r^2$.

Then, for $n=0$, there is a solution of equation (\ref{teste}) in 
$L^2(\mathbb{R}^{+},V^{-1}d\mu)$. Therefore, there is a one-parameter family 
of self-adjoint extensions of $A$ and the spacetime is quantum 
mechanically singular.

Near $r=0$, the negative mass BTZ spacetime is  similar to a
 conic spacetime [see
 equation (\ref{cosmic})]. Positive-frequency solutions 
 of the Klein-Gordon equation in this spacetime satisfies
\begin{equation}
i\frac{\partial \Psi}{\partial t}=(\mu^2-\Delta)^{1/2}\Psi,
\end{equation}
where $\mu$ is the particle mass and $\Delta$ is the Laplace-Beltrami operator
on the cone.

The boundary conditions necessary to
 turn the operator $(\mu^{2} -\Delta_{0})^{1/2}$ self-adjoint, in this case, 
are already known  for a scalar test particle  \cite{kay}.
They are obtained using  Neumann's theory of deficiency  indices (see
 \cite{reed-simon2}) and are given by
\begin{equation}
 \begin{array}{ll}
\lim_{r\to 0}{\big\{\big[\ln(qr/2)+\gamma \big]rR_{0}'(r)-R_{0}\big\}=0},\; &q\in(0,\mu],\\ 
\lim_{r\to 0}{rR_{0}'(r)=0}, &q=0,
\end{array} 
\label{cond}
\end{equation}
where $\gamma$ is Euler-Mascheroni constant. Since $-q^2$ is an 
eingeinvalue of $-\Delta_{0}$, the quantity $q$ is restricted by 
 $0\leq q \leq\mu$ in order that the operator $(\mu^2-\Delta)^{1/2}$ makes sense.
 Note that for  a massless particle ($\mu=0$) we must  take the
 boundary condition for $q=0$. Thus the spacetime is non-singular
 when tested by massless particles.

Because we are interested only in local conditions at 
$r=0$, we take these boundary conditions as the boundary conditions 
of our problem and, given one of the conditions in equation (\ref{cond}), the 
evolution of the wave packet is uniquely determined by the initial data. 
Different choices give us different theories.


\section{DIRAC FIELDS}
\label{SPINOR}

In a $2+1$ dimensional spacetime, fermions have only one spin 
polarization \cite{gavrilov}, hence spinors have only two components 
and the Dirac equation consists of a set of two coupled partial differential 
equations. The constant Dirac matrices, $\gamma^{(j)}$,  in flat
 spacetimes are replaced by the Pauli matrices \cite{sucu}, i.e.,
\begin{equation}
\gamma^{(j)}=(\sigma^{(3)},i\sigma^{(1)},i\sigma^{(2)}),
\end{equation}
where Latin indices represent internal (local) indices. In this way
\begin{equation}
\{\gamma^{(i)},\gamma^{(j)}\}=2\eta^{(ij)}\mathbb{I}_{2\times2},
\end{equation} 
where $\eta^{(ij)}$ is the Minkowski metric in $2+1$ dimensions, 
i.e., $\eta^{(ij)}=\textrm{diag}(-1,1,1)$ 
and $\mathbb{I}_{2\times2}$ is the $2\times 2$ identity matrix.

The coordinate dependent metric $g_{\mu\nu}(x)$ and
 matrices $\sigma^{\mu}(x)$  (Greek indices representing external, or
 global, indices) are related to the  dreibein $e^{(i)}_{\mu}(x)$  by
\begin{equation}
\begin{aligned}
&g_{\mu\nu}(x)=e^{(i)}_{\mu}(x)e^{(j)}_{\nu}(x)\eta_{(ij)},\\
&\sigma^{\mu}(x)=e^{\mu}_{(i)}\gamma^{(i)}.
\end{aligned}
\end{equation}

In the general $2+1$ dimensional spacetime with metric $g_{\mu\nu}(x)$, the
 Dirac equation for a free particle (with mass $M$) can be cast as
\begin{equation}
i\sigma^{\mu}(x)[\partial_{\mu}-\Gamma_{\mu}(x)]\Psi(x)=M\Psi,
\end{equation}
where $\Gamma_{\mu}(x)$ is the spinorial affine connection and it
 is given by
\begin{equation}
\Gamma_{\mu}(x)=\frac{1}{4}g_{\lambda\alpha}
[e^{(i)}_{\nu,\mu}(x)e^{\alpha}_{(i)}(x)
-\Gamma^{\alpha}_{\nu\mu}(x)]s^{\lambda\nu}(x),
\end{equation}
with
\begin{equation}
s^{\lambda \nu}(x)=\frac{1}{2}[\sigma^{\lambda}(x),\sigma^{\nu}(x)].
\end{equation}

As in the case of scalar particles, we are interested in 
the two singular cases ($r\to\infty$ and $r=0$). When $r\to \infty$, 
we take the metric (\ref{infinito}). For this metric, we choose
\begin{equation}
\begin{aligned}
&e^{(i)}_{\mu}(t,r,\theta)=\textrm{diag}(r/l,l/r,r),\\
&e^{\mu}_{(i)}(t,r,\theta)=\textrm{diag}(l/r,r/l,1/r).
\end{aligned}
\end{equation}

The coordinate dependent gamma matrices and the spinorial affine 
connection are given by
\begin{equation}
\begin{aligned}
\sigma^{\mu}(x)&=\bigg(\frac{l}{r}\sigma^{(3)},
\frac{ir}{l}\sigma^{(1)},\frac{i}{r}\sigma^{(2)}\bigg),\\
\Gamma_{\mu}(x)&=\bigg(\frac{1}{2}\frac{r}{l^2}\sigma^{(2)}
,0,\frac{i}{2}\frac{r}{l}\sigma^{(3)}\bigg).
\end{aligned}
\end{equation}

Now, for the spinor
\begin{equation}
\Psi=\left(\begin{array}{lcl}
\psi_{1}\\ 
\psi_2 
\end{array}\right),
\end{equation}
we can write the Dirac equation in the spacetime (\ref{infinito}) as,
\begin{equation}
\begin{aligned}
\frac{il}{r}\frac{\partial \psi_{1}}{\partial t}-
\frac{r}{l}\frac{\partial \psi_{2}}{\partial r}+
\frac{i}{r}\frac{\partial \psi_{2}}{\partial \theta}-
\frac{1}{l}\psi_{2}-M\psi_1&=0,\\
-\frac{il}{r}\frac{\partial \psi_{2}}{\partial t}-
\frac{r}{l}\frac{\partial \psi_{1}}{\partial r}-
\frac{i}{r}\frac{\partial \psi_{1}}{\partial \theta}-
\frac{1}{l}\psi_1-M\psi_2&=0.
\end{aligned} 
\end{equation}

For the positive frequency solutions we shall use the  ansatz,
\begin{equation}
\Psi_{n,E}(t,{\bf x})=\left(\begin{array}{lcl}
R_{1n}(r),\\ 
R_{2n}(r)e^{i\theta} 
\end{array}\right)e^{in\theta}e^{-iEt}.
\end{equation}
Note that $\Psi(t,{\bf x}$) is an eigenfunction of the total 
angular momentum $J_{z}=L_{z}+S_{z}$, with $J_{z}=-
i\frac{\partial}{\partial \theta}$ and $S_{z}=\sigma^{(3)}/2$, with 
eigenvalue $n+\frac{1}{2}$ \cite{kahlilov}. We have for 
the radial part of the Dirac equation,
\begin{equation}
\begin{aligned}
R_{1n}'(r)+\bigg(\frac{1}{r}-\frac{nl}{r^2}\bigg)R_{1n}(r)
+\bigg(\frac{Ml}{r}+\frac{El^2}{r^2}\bigg)R_{2n}(r)&=0,\\
R_{2n}'(r)+\bigg(\frac{1}{r}+\frac{nl}{r^2}\bigg)R_{2n}(r)
+\bigg(\frac{Ml}{r}-\frac{El^2}{r^2}\bigg)R_{1n}(r)&=0.
\end{aligned} 
\label{radial-part}
\end{equation}
By neglecting the lower order terms, since we are interested in the $r\to \infty$
 case, we have:
\begin{equation}
\begin{aligned}
R_{1n}'(r)+\bigg(\frac{1}{r}\bigg)R_{1n}(r)+
\bigg(\frac{Ml}{r}\bigg)R_{2n}(r)&=0,\\
R_{2n}'(r)+\bigg(\frac{1}{r}\bigg)R_{2n}(r)
+\bigg(\frac{Ml}{r}\bigg)R_{1n}(r)&=0.
\end{aligned} 
\label{radial-part2}
\end{equation}

Therefore, for both components we have the same equation,
\begin{equation}
R_{j}''(r)+\frac{3}{r}R_{j}(r)+\frac{1}{r^2}(1
-M^2l^2)R_{j}(r)=0\; (j=1,2).
\end{equation}

Again, neglecting  lower order terms we obtain
\begin{equation}
R_{j}''(r)+\frac{3}{r}R_{j}(r)=0 \qquad(j=1,2).
\label{radial-part3}
\end{equation}

Hence, asymptotically, the radial portion of the 
spinor $\Psi$ behaves as
\begin{equation}
R(r)=Ar^{-2}+B,
\label{radial portion}
\end{equation}
where $A$ and $B$ are constant spinors.

The solution (\ref{radial portion}) is square-integrable only if $B=0$. Only
 one constant must be specified. Then our solution is well-behaved near 
infinity and no extra boundary conditions are necessary.

The metric near $r=0$ is very close to the conic
 background (\ref{cosmic}). This problem has already been 
dealt with in references \cite{gerbert2}  and \cite{gerbert}. The 
metric (after changing the variable $r\to\alpha r$) is,
\begin{equation}
ds^2=-dt^2+\alpha^{-2}dr^2+r^2d\theta^2,
\label{conic modified}
\end{equation}
and the measure on the slice $\Sigma$ is $\alpha^{-1}rdr$.

The appropriate  dreibein is
\begin{align}\label{dreibeinconic}
e^{(i)}_\mu = \left(
              \begin{array}{ccc}
                1 & 0 & 0 \\
                0 & \alpha^{-1}\cos{\theta} & \alpha^{-1}\sin{\theta} \\
                0 & -r\sin{\theta} & r\cos{\theta}
              \end{array}
             \right).
\end{align}
And by choosing positive energy solutions of the form,
\begin{equation}
\Psi_{n,E}(t,{\bf x})=\left(\begin{array}{ccc}
R_{1n}(r),\\ 
iR_{2n}(r) 
\end{array}\right)e^{i(n+\frac{1}{2}-\frac{1}{2}\sigma^{(3)})\theta}e^{-iEt},
\end{equation}
we obtain the system of equations,
\begin{equation}
\begin{aligned}
R_{1n}'(r)+\bigg[\frac{1}{2r}-\frac{n+\frac{1}{2}}{\alpha 
r}\bigg]R_{1n}(r)+\frac{E+M}{\alpha}R_{2n}(r)&=0,\\
R_{2n}'(r)+\bigg[\frac{1}{2r}+\frac{n+\frac{1}{2}}{\alpha
 r}\bigg]R_{2n}(r)-\frac{E-M}{\alpha}R_{1n}(r)&=0.
\end{aligned}
\label{radial-dirac}
\end{equation}

A complete set of solutions of the Dirac equation in the 
spacetime (\ref{conic modified}) is given by the normal modes,
\begin{widetext}
\begin{equation}
\Psi_{n,E}(t,{\bf x})=\bigg[A_{n}\left(\begin{array}{ccc}
J_{|\nu|}(\kappa r)\\ 
iJ_{|\nu+1|}(\kappa r) 
\end{array}\right) +B_{n}\left(\begin{array}{ccc}
N_{|\nu|}(\kappa r)\\ 
-iN_{|\nu+1|}(\kappa r) 
\end{array}\right)\bigg]e^{i(n+\frac{1}{2}
-\frac{1}{2}\sigma^{(3)})\theta}e^{-iEt},
\label{solution2}
\end{equation}
\end{widetext}
where $\nu=[2n+(1-\alpha)]/2\alpha$, 
$\kappa^2=(E^2-M^2)/\alpha^2$,  $A_{n}$ and $B_{n}$ are arbitrary
 constant. The 
 Bessel function $J_{\lambda}$, $\lambda \in \mathbb{R}$, 
is square-integrable for all $\lambda$, but the Neumann 
function $N_{|\lambda|}$ is not, except when $|\lambda|<1$. Note
 that the second spinor in (\ref{solution2}) is square-integrable 
when $|\nu|<1$ and $|\nu+1|<1$, i.e., $-1<\nu<0$. It is easy to see
 that this condition does not hold for any value of $n\in\mathbb{Z}$. 
Therefore, an arbitrary wave packet can be described by
\begin{equation}
\Psi(t,{\bf x})=\sum_{n=-\infty}^{+\infty}{A_{n}\left(\begin{array}{ccc}
J_{|\nu(n)|}(\kappa r)\\ 
iJ_{|\nu(n)+1|}(\kappa r) 
\end{array}\right)}
e^{i(n+\frac{1}{2}
-\frac{1}{2}\sigma^{(3)})\theta}e^{-iEt}
\label{solution3}
\end{equation}
and the initial condition $\Psi(0,\bf x)$ is sufficient  to determine the time evolution
of the particle. The Cauchy problem is well-posed and the spacetime is nonsingular when 
tested by fermions. It is interesting to note that this is the case only for the  $2+1$ 
dimensional spacetime, since for the  $3+1$ dimensional case (see \cite{KONKOWSKI2}), 
the extra dimension adds a continuous parameter $k$ representing the wave vector in the 
Fourier transform of $\Psi$. The existence of this  continuous parameter  allows 
 that an infinite number  of normal modes to  be singular. So  the spacetime around 
a cosmic string remains singular when tested by fermions. The fact that in
 the  $2+1$ dimensional spacetime the spatial 
modes have only discrete indices excludes this possibility.

\section{CONCLUSION}
\label{CONCLUSION}

The BTZ spacetime with mass parameter $m\geq 0$ does not cause any problem since 
the weak cosmic censorship hypothesis  is satisfied (the singularity
 is hidden). For 
the mass parameter interval $-1<m<0$, the classical singularity 
persists when tested by massive scalar fields. In principle, we do not have any 
reason to choose one  of the boundary conditions in (\ref{cond}). As well as, it is
 uncertain the future of a  classical particle moving along a  geodesic which 
reaches the singularity after a finite  time, it is  uncertain the time evolution
of the corresponding quantum particle, provided it obeys the Klein-Gordon 
equation. But, when tested by massless scalar bosons and by fermions, the singularity is ``healed'' 
and no extra boundary conditions are  necessary. The spacetime is wave regular for these fields.

Because the BTZ spacetime with negative mass parameter is not regular for every 
quantum particle, it must be excluded by the weak cosmic censorship hypothesis. For this reason, 
the $m=-1$ case, which is the stable ground state of the  BTZ spacetime when studied in the  general 
relativity context, remains stable when studied in the  quantum mechanics framework.
\acknowledgements 
We thank CNPq for financial  support and  P.S.L. also thanks  FAPESP.




\begin{thebibliography}{99}

\bibitem{jackiw1}
 S. Deser, R. Jackiw  and G. t'Hooft,  Ann. Phys. {\bf 152}, 220 (1984)
 
\bibitem{staruszkiewicz}
A. Staruszkiewicz, 
 Acta. Phys. Polon. {\bf 24}, 734 (1963).
 
\bibitem{letelier}
 P.S. Letelier, 
 Class. Quantum Grav. {\bf 4}, 75 (1987).

\bibitem{konkowski1}
D.A. Konkowski  and T.M. Helliwell, 
 Gen. Relativ. Grav.  {\bf 33}, 1131 (2001).
 
\bibitem{btz1}
M. Ba\~nados, C. Teitelboim and J. Zanelli,  
Phys. Rev. Lett.  {\bf 69},  1849 (1992). 

\bibitem{btz2}
M. Ba\~nados,   M. Henneaux,  C. Teitelboim  and J.  Zanelli,
 Phys. Rev. D  {\bf 48}, 1506 (1993).

\bibitem{jackiw2}
 S. Deser  and R.  Jackiw,  Ann. Phys. {\bf 153}, 404.(1984).

\bibitem{pantoja}
 N. Pantoja, H. Rago  and  R.  Rodr\'\i guez,
J. Math. Phys. {\bf 45},  1994 (2004). 

\bibitem{horowitz2}
G.T. Horowitz  and R. Myers,  
 Gen. Relativ. Gravit. {\bf 27},  915 (1995).

\bibitem{horowitz}
G.T. Horowitz  and  D. Marolf, 
Phys. Rev. {\bf 52}, 5670 (1995). 

\bibitem{kay}
B.S. Kay  and  U.M. Studer,  
Comm. Math. Phys. {\bf 138}, 103 (1991). 

\bibitem{gerbert2}
 P. de Sousa Gerbert,
Phys. Rev. D {\bf 40}, 1346 (1989).

\bibitem{gerbert}
P. S. Gerbert and R. Jackiw,
Commun. Math. Phys. {\bf 124}, 229 (1989).


\bibitem{hawking}
S.  Hawking and G. Ellis,   
 {\it The Large Scale Structure of Space-Time},
(Cambridge University Press, Cambridge, 1973 ).

\bibitem{reed-simon1}
M.  Reed and B. Simon,   
 {\it Functional Analysis},
(Academic Press, New York, 1980 ).

\bibitem{reed-simon2}
M. Reed  and B. Simon,   
 {\it Fourier Analysis and Self-Adjointness},
(Academic Press, New York,  1975 ).

\bibitem{richtmyer}
R.D.  Richtmyer, 
{\it Principles of Advanced Mathematical Physics},
(Springer, New York, 1978).

\bibitem{gavrilov}
S.P. Gavirilov, D.M. Gitman and J.L. Tomazelli,
Eur. Phys. J. C {\bf 39}, 245 (2005).

\bibitem{sucu}
 Y. Sucu and Nuri \"Unai,
J. Math. Phys. {\bf 48}, 052503 (2007).

\bibitem{kahlilov}
 V.R. Kahlilov and C.L. Ho,
Mod. Phys. Lett. A {\bf 13}, 615 (1998).

\bibitem{KONKOWSKI2}
T.M. Helliwell, D.A. Konkowski and V. Arndt 
Gen. Rel. Grav. {\bf 35}, 79 (2003). 



\end{thebibliography}
\end{document}